\documentclass[aps,prl,twocolumn,superscriptaddress,groupedaddress]{revtex4}  % for review and submission
\usepackage{graphicx}  % needed for figures
\usepackage{dcolumn}   % needed for some tables
\usepackage{bm}        % for math
\usepackage{amssymb}   % for math
\usepackage[]{amsmath,amssymb}
\usepackage{graphics,epsfig}
\begin{document}

%\input author_list.tex       % D0 authors (remove the first 3 lines
                             % of this file prior to submission, they
                             % contain a time stamp for the authorlist)
                             % (includes institutions and visitors)
%\date{\today}

\title{ Localization of negative energy and the Bekenstein bound}
%\title{Bounds on negative energy distribution in CFT}
\author{ David D. Blanco and Horacio Casini\\
{\sl Centro At\'omico Bariloche,
8400-S.C. de Bariloche, R\'{\i}o Negro, Argentina}}

\begin{abstract}
A simple argument shows negative energy cannot be isolated far away from positive energy in a conformal field theory, and strongly constrains its possible dispersal. This is also required by consistency with Bekenstein bound written in terms of positivity of relative entropy. We prove a new form of Bekenstein bound based on monotonicity of relative entropy, involving a ``free'' entropy enclosed in a region which is highly insensitive to space-time entanglement, and show it further improves the negative energy localization bound.   
\end{abstract}

\maketitle

{\sl Introduction.---} Due to Lorentz symmetry and the existence of a fundamental state, energy is always positive in quantum field theory (QFT) \cite{pct}. However, energy density can take negative values if it is compensated by the presence of positive energy in other regions of the space. In fact, in any QFT there are necessarily some states having negative energy density \cite{nega-necesaria}. This is a purely quantum phenomenon which in general is not expected to survive the classical limit. For example, the classical energy density for a free scalar field
$T^{00}(x)=\frac{1}{2}\left(\dot{\phi}^2+(\nabla \phi)^2+m^2 \phi^2\right)$ 
is positive definite. In the process of quantization the subtraction of zero point energy renders the energy density operator indefinite.

Various energy conditions, stating generically the positivity of some combinations of the stress tensor components, have found important applications in classical gravity. For example, strong energy condition is related to the singularity theorem for cosmological solutions \cite{cosmo}, and null energy condition is an assumption of Hawking's area theorem, implying black hole horizon area increases with time \cite{area}. Quantum violation of the null energy condition is then necessary for black hole evaporation. Enough negative energy is also required for the existence of traversable wormholes and time-machines \cite{wormhole}. More recently, in the context of holographic models, energy conditions on a classical bulk space-time have been related to properties on the boundary QFT such as strong subadditivity of entanglement entropy \cite{wall} and the renormalization group irreversibility (c-theorem) \cite{cteo}.

In connection with these applications, it is of great interest to know how much negative energy density quantum mechanics can support in violation of the classical energy conditions. While an answer to this question in general curved space is out of sight, some important progress have been made in Minkowski space \cite{review}. 

A {\sl quantum energy inequality} is generically a bound on a combination of expectation values of the stress tensor components weighted by some space-time function. 
Several of these bounds have been worked out in the literature \cite{varios} (see also \cite{review} and references therein). However, most of the known examples apply only to free fields, and typical quantum energy inequalities  do not constrain the spatial distribution of negative energy but assume the form of a bound on the possible duration in time of the negative energy density at a specific point in space (see, however, \cite{oso}). We show below that a simple argument for conformal field theories (CFT) provides us with a generic constraint on the spatial distribution of negative energy. More precisely, negative energy appears to be confined to live near positive energy and has to be less disperse than positive energy.

As argued by Ford \cite{ford}, constraints on the availability and manipulation of negative energy are necessary for the validity of the second law of thermodynamics. For example, dropping negative energy on a black hole could reduce its size and entropy without a compensation in the emitted entropy through increased Hawking radiation.

 Interestingly, a thought experiment involving black holes and the generalized second law (GSL) also gives place to the Bekenstein bound \cite{beke},
 \begin{equation}
S_A\le 2 \pi \,R \,E_A\,,\label{bek}
\end{equation} 
where $S_A$ and $E_A$ are the entropy and energy of any object which can be enclosed in a region $A$ of circumscribing radius $R$. Since quantum mechanical entropy is positive, this would mean that energy contained in a region cannot be negative. Of course, this is not strictly correct, and the reason is that the quantities involved in (\ref{bek}) have to be defined with some care in QFT. For example, entanglement entropy of vacuum fluctuations across the boundary gives an infinite contribution to the bare entropy of a region. A naive interpretation of (\ref{bek}) also seems to 
 indicate there should be a bound on the number of field species, while this is not implied by the GSL, as was widely discussed in the literature \cite{ms,dis,beke1}.  

A well defined quantum version of Bekenstein bound requires we write the left hand side of (\ref{bek}) as a subtraction $\Delta S_A=S_A^1-S_A^0$ between the entropy $S_A^1=-\textrm{tr} \rho_A^1 \log \rho_A^1$ of the state of the object $\rho_A^1$ reduced to region $A$ and the entropy of the vacuum state $S_A^0=-\textrm{tr}\rho_A^0 \log \rho_A^0$ in the same region \cite{beke1,beke2}.  This eliminates the ultraviolet divergent terms of the entropy which are artificially produced by localization, and also solves the species problem \cite{ms,beke1}.  Additionally, the product $2\pi E R$ on the right hand side of (\ref{bek}) has to be written in terms of the {\sl modular Hamiltonian} $H_A=-\log (\rho_A^0)$ corresponding to the reduced density matrix of the vacuum in $A$. The relation between $H_A$ with energy and size is clarified if we take $A$ to be the half space $x^1>0$. In this case $H_A$ is given by the generator of boost symmetry inside $A$, for any QFT and space-time dimension $d$, 
\begin{equation}
H_A=2\pi \int_{x^1>0} d^{d-1}x \,\,x^1 \,T^{00}(x)\,.\label{boost}  
\end{equation}
Hence, taking into account expression (\ref{boost}), a natural quantum interpretation of (\ref{bek}) reads \cite{beke2}
\begin{equation}
\Delta S_A\le \Delta \langle H_A\rangle\,,\label{ine}
\end{equation}
where $\Delta \langle H_A\rangle= \textrm{tr}(\rho_A^1 H_A)-\textrm{tr}(\rho_A^0 H_A)$ is the variation of the expectation value of the modular Hamiltonian between the object state and the vacuum state. 
  In the form (\ref{ine}), the bound is valid for any region $A$ (not necessarily the half space) and for any ``object'' state $\rho^1$, due to the positivity of {\sl relative entropy} 
 \begin{equation}
 S(\rho_A^1|\rho_A^0)=\textrm{tr}(\rho^1_A \log \rho^1_A-\rho^1_A \log \rho^0_A)=\Delta \langle H_A\rangle-\Delta S_A\ge 0
 \end{equation}
  between the object's state and vacuum state, both reduced to $A$. Relative entropy $S(\rho^1|\rho^0)$ is a central quantity in quantum information theory, which essentially measures distinguishability between two states.

The bound (\ref{ine}) is free from divergences and holds universally. Its validity depends on quantum mechanics and relativity, and therefore it is not a new constraint for flat space physics coming from black holes, as was originally thought.   

Eq. (\ref{ine}) allows (in contrast to the naive interpretation of (\ref{bek})) for negative values on both sides of the inequality. 
This is because the expectation value of (\ref{boost}) can be negative for some states. In this case, negative energy in $A$ must be accompanied by a decrease of the entanglement entropy of the state with respect to the entanglement entropy of the vacuum. 

However, consistency of the inequality (\ref{ine}) and the one corresponding to the complementary region $\bar{A}$, 
requires some constraint on the distribution of negative energy, as was recently suggested \cite{relative}. Following this idea,  we 
find here a new quantum version of Bekenstein bound (i.e. different from (\ref{ine})) which is also universally valid and involves only positive quantities on both sides of the inequality. This form of Bekenstein bound improves our bound on negative energy localization coming purely from conformal symmetry arguments.

{\sl A positive symmetry generator.---}
In a Lorentz covariant theory we can unitarily transform the Hamiltonian with a boost to any operator of the form $P_\mu a^\mu$, with $a^\mu$ a vector in the future light cone. This immediately tell us these operators are positive definite, since they have the same spectrum as the Hamiltonian. In a CFT Lorentz group is part of a larger group of conformal transformations, which move the Hamiltonian in a larger cone of positive operators. 

Then, let us make a conformal transformation of the Hamiltonian. To keep the result as much symmetric as possible, consider first transforming with the conformal transformation $\hat{I}=R.I$, where $R$ is a spatial reflection and $I$ is the coordinate inversion $x^{\mu\prime}=\frac{x^\mu}{x^2}$. 
The spatial reflection $R$ is necessary to make $R.I$ belong to the conformal group connected to the identity. 
The composite coordinate transformation $\hat{I}^{-1}. \delta t . \hat{I}$, where $\delta t$ is a time translation of small amount $\delta t^\mu\equiv (\delta t,0,0,0)$ is, to first order in $\delta t$,
\begin{equation}
x^{\mu\prime}\simeq x^\mu+  \, x^2 \, \delta t^\mu - 2  x^\mu (\delta t^\alpha x_\alpha)\,.
\end{equation}
We can read off the generator $G$ implementing this conformal transformation looking at the effect on the points of the surface $x^0=0$,
 \begin{equation}
G=\int d^{d-1}x\, |\vec{x}|^2 T^{00}(x)\,.\label{this}
\end{equation}
 Hence, we have for the quantum operators $G=\hat{I}^\dagger.H.\hat{I}$, and $G$ is positive definite.

A straightforward examination of the general form of a conformal generator shows the most general one written only in terms of the energy density (i.e. not involving the momentum density $T^{0i}$) that we can get from a conformal transformation of $P^0$ is a linear combination with positive coefficients of the Hamiltonian and the translates of $G$. 

{\sl Negative energy localization.---} 
The positivity of $G$ means a ``moment of inertia'' of the energy density is positive. 
We have for the expectation values in any state 
\begin{equation}
\int d^{d-1}x\,\, |\vec{x}-\vec{x}_0|^2 \langle T^{00}(x)\rangle \ge 0\,.\label{30}
\end{equation}
 The most constraining bound follows minimizing (\ref{30}) over the position of $\vec{x}_0$.

In order to clarify the meaning of (\ref{30}) for the energy distribution let us call the total positive energy $E_+$ and the absolute value of the negative energy $E_-$,
\begin{equation}
E_\pm=\int d^{d-1}x\,\,\theta(\pm\langle T^{00}(x)\rangle)\,\,|\langle T^{00}(x)\rangle|\,.
\end{equation}  
Then, we have for the total energy $E=E_+-E_-\ge 0$. We also define the positive and negative energies center of mass $\vec{x}_\pm$ as
\begin{equation}
\vec{x}_\pm E_\pm=\int d^{d-1}x\, \vec{x} \,\theta(\pm\langle T^{00}(x)\rangle)\,\,|\langle T^{00}(x)\rangle|\,,
\end{equation}
and the mean square size $r_\pm$ of the positive and negative distributions as
\begin{equation}
(r_\pm)^2 E_\pm =\int d^{d-1}x\, |\vec{x}-\vec{x}_{\pm}|^2 \,\theta(\pm\langle T^{00}(x)\rangle)\,\,|\langle T^{00}(x)\rangle|\,.
\end{equation}  

Then, taking into account that the $\vec{x}_0$ which minimizes (\ref{30}) is
\begin{equation}
\vec{x}_0=\frac{E_+ \vec{x}_+-E_- \vec{x}_-}{E}\,,\label{punto}
\end{equation}
we get the following bound
\begin{equation}
|\vec{x}_+-\vec{x}_-|^2\le E\frac{E_+ r_+^2-E_- r_-^2}{E_+ E_-}\,.\label{bi}
\end{equation}
In particular, the intrinsic size of negative energy ``moment of inertia'' is bounded by the one of positive energy 
\begin{equation}
 E_- r_-^2\le E_+ r_+^2 \,.
\end{equation}
From (\ref{bi})  positive and negative energies cannot be separated too much. For example, if we take a small negative energy density region $r_-\ll r_+$, and a small amount of negative energy, $E_-\ll E_+$, we have $|\vec{x}_+-\vec{x}_-|\lesssim \sqrt{\frac{E}{E_-}} r_+$. 

{\sl Positivity from relative entropy.---} 
The connection of the positivity of $G$ with relative entropy comes through the fact that in CFT the modular Hamiltonian for a spherical region in the vacuum state is proportional to the generator of conformal transformations that keep the sphere fixed \cite{modular}. The modular Hamiltonian for a sphere $A$ of radius $R$ is   
\begin{equation}
H_A=2 \pi \int_{|\vec{x}|\le R} d^{d-1}x\, \frac{R^2-|\vec{x}|^2}{2 R} T^{00}(x)\,,\label{tera}
\end{equation}
while for the complementary region $\bar{A}$ (the space outside $A$) it is
\begin{equation}
H_{\bar{A}}=2 \pi \int_{|\vec{x}|\ge R} d^{d-1}x\, \frac{|\vec{x}|^2-R^2}{2 R} T^{00}(x)\,.\label{tira}
\end{equation}

For the sake of the argument, let us first think in more general terms, and take an arbitrary region $A$ in a general QFT, not 
necessarily a conformal theory. Let us call the operator 
\begin{equation}
\hat{H}_A=H_A-H_{\bar{A}}=-\log(\rho^0_A)\otimes 1+1\otimes \log{\rho^0_{\bar{A}}}\label{full}
\end{equation}
 the full modular Hamiltonian of $A$. Writing the vacuum state in Schmidt decomposition across the tensor product ${\cal H}_A\otimes{\cal H}_{\bar{A}}$, a direct calculation shows 
\begin{equation}
\hat{H}_A|0\rangle=(H_A-H_{\bar{A}})|0\rangle=0\,.\label{refi}
\end{equation}

Let us now consider another global state $\rho^1$ different from the vacuum and also a smaller region $B\subseteq A$. The relative entropy between $\rho^1$ and $\rho^0$ in a region $X$ is $S(\rho_X^1|\rho_X^0)=\Delta \langle H_X \rangle- \Delta S_X$, and it is both positive and monotonically increasing with the region size \cite{wehrl}. From monotonicity we have 
\begin{equation}
\Delta \langle H_A\rangle- \Delta S_A\ge \Delta \langle H_B\rangle- \Delta S_B\,,\label{ii}
\end{equation}
and also
\begin{equation}
\Delta \langle H_{\bar{B}}\rangle- \Delta S_{\bar{B}}\ge \Delta \langle H_{\bar{A}}\rangle- \Delta S_{\bar{A}}\,.\label{jj}
\end{equation}

Property (\ref{refi}) gives $\langle H_A\rangle^0 =\langle H_{\bar{A}}\rangle^0$ and $\langle H_B\rangle^0 =\langle H_{\bar{B}}\rangle^0$. Since the vacuum state is pure, we also have $ S_A^0=S_{\bar{A}}^0$ and  $S_B^0= S_{\bar{B}}^0$. Then, adding (\ref{ii}) and (\ref{jj}) we get 
\begin{equation}
\langle \hat{H}_A-\hat{H}_B\rangle^1 \ge S_A^1-S_B^1+S_{\bar{B}}^{1}-S_{\bar{A}}^1\equiv 2\, S_f(A,B)\,.\label{perio}
\end{equation}
In this inequality the vacuum is present only through the definition of the modular Hamiltonians. The combination of entropies in the right hand side, which for later convenience we have called $2\, S_f(A,B)$, is always positive as a consequence of weak monotonicity property, $S(X)+S(Y)\ge S(X-Y)+S(Y-X)$ applied to $X=A$, $Y=\bar{B}$ \cite{ffff}. This inequality is in turn a well known direct consequence of strong subadditivity of the entropy. Hence, as (\ref{perio}) is valid for any $\rho^1$, the difference $\hat{H}_{A}-\hat{H}_B$ for $B\subseteq A$ is a positive operator \cite{1}.  

Coming back to the case of spheres in a CFT, we can choose $A$ to be a sphere of radius $R_1$ and $B$ a concentric smaller sphere of radius $R_2$, with $R_2<R_1$. Using (\ref{tera}) and (\ref{tira}) we get 
\begin{equation}
\frac{\pi}{2} (R_1-R_2)\left(\langle P^0\rangle +  \langle G\rangle/(R_1 R_2)\right)\ge \,S_f(A,B)>0\,.\label{poro}
\end{equation}
Taking the limit $R_2\rightarrow 0$ we recover the positivity of $G$ in equation (\ref{this}).

{\sl A new quantum Bekenstein bound.---} 
Inequality (\ref{perio}) is our proposal for a new, universally valid, quantum Bekenstein bound. To see how this compares with the original formulation (\ref{bek}), let us apply  (\ref{perio}) to the case of two half-spaces included into one another, i.e. $A$ is the region $x^1>0$, $B$ is given by $x^1>L>0$, and the region $A-B$ is a strip of width $L$. Using (\ref{boost}) we get
\begin{eqnarray}
 \pi\, L\,  E\ge S_f(A,B)=\label{piso}\frac{1}{2}( S(x^1>0)-S(x^1>L)\nonumber \\
 +S(x^1<L)-S(x^1<0))\,.
\end{eqnarray}

 In the classical limit, this strongly resembles Bekenstein original formulation (\ref{bek}). 
 To have a feeling of the entropic quantity $S_f(A,B)$ on the right hand side, one can imagine evaluating it for a thermal gas at high temperature. Then, while most of the contribution of the entanglement around the boundaries cancel out in the combination, $S_f(A,B)$ will capture exactly the extensive entropy of the gas inside $A$ but outside $B$. This is the reason we have inserted a factor $2$ in our definition of $S_f(A,B)$ in (\ref{perio}). Note also that for a pure global state $S(X)=S(\bar{X})$, and $S_f(A,B)\equiv 0$. Then, $S_f(A,B)$ does not capture the entropy in the strip $A-B$ produced by an entangled pair of particles, one of which is in $A-B$ and the other is outside it. Hence, we are lead to interpret the quantity $S_f(A,B)$, to a certain extent, as a {\sl ``free'' (or ``global'') entropy located in between the boundaries of $A$ and $B$}. 
 
 In this sense it is  clarifying to write the entropy of the global state as coming from partial tracing over a hidden sector $\aleph$ which is used to purify it, i.e $\rho^1=\textrm{tr}_{\aleph} |\psi\rangle \langle\psi|$, for some vector $|\psi\rangle$ in ${\cal H}\otimes \aleph$. Using this representation we can write 
 \begin{equation}
 S_f(A,B)=\frac{I(A,\aleph)-I(B,\aleph)}{2}\,,
 \end{equation}           
where $I(X,Y)=S(X)+S(Y)-S(X\cup Y)$ is the mutual information between $X$ and $Y$. Hence, $S_f(A,B)$ is approximately extensive and depends only on $A-B$ to the extent that $I(X,\aleph)$ is approximately extensive for spatial regions $X$. This representation also shows $S_f(A,B)$ is in fact monotonically increasing with the size of $A-B$ because mutual information is a monotonically increasing quantity. Moreover, it trivially satisfies a partial form of extensivity: If $C\subset B$ we have $S_f(A,C)=S_f(A,B)+S_f(B,C)$.
 
While the general aspect of (\ref{piso}) is similar to Bekenstein's original formulation, there are also some interesting differences.
For example, the strip-like  region has now $L$ as a {\sl minimal} size of the region rather than the circumscribing diameter in the original formulation of the Bekenstein bound (in this respect our bound is similar to the proposal of \cite{bousso}). Also, in the left hand side we have now the global energy $E$ instead of a measure of the energy in a region.

The new formulation (\ref{piso}) (or more generally (\ref{perio})) has also some remarkable differences with respect to the first quantum version (\ref{ine}) of Bekenstein bound. First, (\ref{piso}) is about energy and entropies in one state, rather than the difference between two states as in (\ref{ine}). Mathematically, (\ref{piso}) comes from a very different inequality: Monotonicity of relative entropy, rather than from positivity of relative entropy as it was the case of (\ref{ine}).  Finally, in contrast to (\ref{ine}), both sides of the new inequality are now positive. 
 
\begin{figure}
\centering
\leavevmode
\epsfysize=3cm
\epsfbox{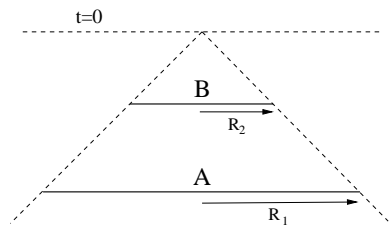}
\bigskip
\caption{Two spatial spheres $A$ of radius $R_1$ and $B$ of radius $R_2$ located on the past light cone. $S_f(A,B)$ is a measure of entropy crossing the null cone in between the boundaries of $A$ and $B$.}
\label{ffff}
\end{figure}

 {\sl Entropy and negative energy distribution.---} 
We now use again the inequality (\ref{perio}) for two spheres in a CFT, but since we want to have a bound in terms of the operator $G$ alone, and not containing a contribution of the Hamiltonian as in eq. (\ref{poro}), we are forced to use spheres located at different times.  We take as $A$ a sphere of radius $R_1$ at time $t=-R_1$, centered at the spatial origin, and as $B$ another sphere of radius $R_2<R_1$ centered at the origin but lying at time $t=-R_2$ (see figure \ref{ffff}). Hence, $A$ and $B$ lie in the past light cone.

The conformal current that gives the conformal generator corresponding to $\hat{H}_A$ is 
\begin{equation}
J_A^\mu
= 2 \pi T^{\mu\beta} x_\beta+T^{\mu\beta}\left(c^\alpha x_\alpha x_\beta-\frac{1}{2} c_\beta x^\alpha x_\alpha\right)
\end{equation} 
with $c_\beta\equiv(2 \pi/R_1,0,...,0)$. 
The current $J_B^\mu$ follows from $J^\mu_A$ by replacing $R_1$ by $R_2$. Hence, in this case the bound is written
\begin{eqnarray}
&&\langle \hat{H}_A-\hat{H}_B\rangle=\int d^{d-1}x\, (J^0_A-J^0_B)\\&&=\pi \left(\frac{1}{R_2}-\frac{1}{R_1}\right) \int |\vec{x}|^2\,\langle  T^{00}(x)\rangle \ge 2\,S_f(A,B)\,.\nonumber
\end{eqnarray} 
In particular, choosing the center of spatial coordinates at the point $\vec{x}_0$ in (\ref{punto}), we get
\begin{eqnarray}
&&(E_+ r_+^2-E_- r_-^2)-\frac{E_+ E_-}{E}|\vec{x}_+-\vec{x}_-|^2\nonumber\\ &&\hspace{.4cm}\ge\max_{R_1,R_2,R_1>R_2}\left(\frac{2 R_1 R_2}{(R_1-R_2)\pi}\,S_f(R_1,R_2)\right)\,.
\end{eqnarray}

Therefore, entropy makes the localization bound more restrictive. This is quite natural from the point of view of the original motivation on negative energy bounds based on the second law \cite{ford}: A pure state with negative energy which merges with a thermal state decreases its energy, and consequently the phase space available, possibly reducing the entropy and violating the second law. In this sense, matter with negative energy but positive entropy could only worsen the problem, since the second law would be violated by a larger amount. However, entanglement entropy between the negative energy source and the positive energy reservoir is clearly not an additional problem, since this entropy disappears once they have merged; in other terms, this entanglement entropy is not considered in the balance of entropy for the global initial and final states in the second law. This is reflected in that the specific entropic quantity that enters the bound does not feel spatial entanglement. 

As a final comment, the existence of a quantum Bekenstein bound based on monotonicity of relative entropy suggests this property is important for the validity of the GSL. In fact such connection has been pointed out in the literature \cite{aron}. 

{\sl Acknowledgements.---} H.C. thanks correspondence with Juan Martin Maldacena and Raphael Bousso which stimulated the initiation of this work. This work was supported by CONICET, Universidad Nacional de Cuyo, and CNEA, Argentina.

\end{document}